\documentclass{aastex701}

\newcommand{\DM}{\mathrm{DM}}
\newcommand{\RM}{\mathrm{RM}}
\usepackage{amsmath}

\begin{document}


\title{
Probing Solar-Wind Structure During Crab Pulsar Conjunctions
with Dispersion and Faraday Rotation
}


\author{Huiyu Xiao}
\affiliation{
Department of Physics,
University of Toronto,
Toronto, ON, Canada
}
\email{huiyu.xiao@mail.utoronto.ca}

\author{Daniel Baker}
\affiliation{
Academia Sinica Institute of Astronomy and Astrophysics,
Taipei, Taiwan
}
\email{dbaker@asiaa.sinica.edu.tw}

\author{Thierry Serafin Nadeau}
\affiliation{
David A. Dunlap Department of Astronomy and Astrophysics,
University of Toronto,
Toronto, ON, Canada
}
\affiliation{
Canadian Institute for Theoretical Astrophysics,
University of Toronto,
Toronto, ON, Canada
}
\email{thierry.serafinnadeau@mail.utoronto.ca}

\author{Hao Shen}
\affiliation{
University of Science and Technology of China,
Hefei, Anhui, China
}
\email{haoshenustc@mail.ustc.edu.cn}

\author{Dongzi Li}
\affiliation{
Tsinghua University,
Beijing, China
}
\email{dzli@tsinghua.edu.cn}
\author{Ue-Li Pen}
\affiliation{
Academia Sinica Institute of Astronomy and Astrophysics,
Taipei, Taiwan
}
\affiliation{
Canadian Institute for Theoretical Astrophysics,
University of Toronto,
Toronto, ON, Canada
}
\email{pen@asiaa.sinica.edu.tw}


\begin{abstract}
Observations of the Crab pulsar (PSR B0531+21) during solar
conjunction probe plasma structure in the inner solar wind through
radio-propagation effects. We analyze dispersion measure (DM) and
rotation measure (RM) measurements from the annual conjunctions of
2022--2026, obtained with the Canadian Hydrogen Intensity Mapping
Experiment (CHIME). The
2025 DM time series is distinguished from those of the other years by
a secondary enhancement on 21 June, about one week after the main
conjunction peak. In our scattering-corrected DM measurements, this
feature, $\Delta\DM=(3.95\pm0.20)\times10^{-3}~\mathrm{pc\,cm^{-3}}$,
is about 20 times the day-to-day DM scatter outside the conjunction
window. The RM time series, measured by combining the single pulses of
each observation and corrected for the ionospheric contribution, show
conjunction excursions of up to $\sim60~\mathrm{rad\,m^{-2}}$ and
changes of sign consistent with crossings of heliospheric
magnetic-sector boundaries, but no counterpart to the secondary 2025
DM feature or to a similar post-conjunction DM tail in 2026. From the
slope of the RM excess against the DM excess we estimate the
electron-density-weighted line-of-sight magnetic field; for the
sectors with significant slopes, $|\langle B_\parallel\rangle|$ is at
least $\sim0.2$ to $\sim10$~mG near conjunction, largest in 2025.
\end{abstract}



\section{Introduction}
\label{sec:introduction}

The solar wind is a continuously expanding plasma originating from
the solar atmosphere, and its density and magnetic-field structure
are fundamental properties of the heliospheric environment.
Radio observations of compact astrophysical sources provide a
sensitive means of probing this plasma remotely, particularly when
the source lies at a small angular separation from the Sun.

Radio signals propagating through ionized plasma can be modified by
dispersion, Faraday rotation, scattering, and scintillation. These
effects provide complementary information about the electron density
and magnetic field along the line of sight
\citep{You2012,Kumar2022}. Pulsars are particularly useful for such measurements because their
short-duration radio pulses allow propagation-induced effects to be
measured precisely. 

The dispersion measure is defined as

\begin{equation}
    \DM
    =
    \int_{\mathrm{LOS}} n_e\,dl,
    \label{eq:dm}
\end{equation}

where $n_e$ is the free-electron number density along the line of
sight. Physically, dispersion arises because the refractive index of a
cold ionized plasma depends on radio frequency, causing lower-frequency
pulse components to arrive later than higher-frequency components; the
relative delay scales approximately as $\DM\,\nu^{-2}$.

Faraday rotation occurs because the two circularly polarized modes
propagate at different phase velocities through a magnetized plasma,
rotating the plane of linear polarization. The polarization position
angle varies as $\psi(\lambda)=\psi_0+\RM\lambda^2$. The rotation
measure is given by

\begin{equation}
    \RM
    =
    0.812
    \int_{\mathrm{LOS}}
    n_e B_{\parallel}\,dl,
    \label{eq:rm}
\end{equation}

for $n_e$ in $\mathrm{cm^{-3}}$, $B_{\parallel}$ in $\mu\mathrm{G}$,
and $dl$ in pc, giving RM in $\mathrm{rad\,m^{-2}}$. Here
$B_{\parallel}$ is the line-of-sight magnetic-field component. Simultaneous
measurements of DM and RM can therefore provide complementary
constraints on electron-density and magnetic-field variations
\citep{You2012}.

Previous studies have measured time-dependent DM and RM variations
associated with the solar wind. In particular, observations of
giant pulses from the Crab pulsar have been used to measure
coronal-density variations near solar conjunction
\citep{Tokumaru2022SoPh}. Those observations were conducted near
327 MHz, lower than the 400--800 MHz observing band considered in
the present work. They found excess DM above the interstellar background during the
2018 and 2019 conjunctions and fitted a spherically symmetric coronal
electron-density model to the measurements. The inferred density
profiles were relatively flat compared with an $r^{-2}$ radial law,
consistent with the influence of a southern polar coronal hole during
the solar-cycle 24/25 minimum. Faraday rotation measurements of PSR0525 + 21 during solar occultation  have also been used to probe magnetic fields in the solar corona \citep{Bird1980}.

Pulsar-timing studies have developed increasingly flexible models of
the solar-wind DM. The simplest model is a spherically symmetric
$n_e\propto r^{-2}$ wind with a single density at 1~au, which
\citet{Madison2019} fit across the NANOGrav 11-year data set.
\citet{You2007} instead separated slow and fast wind using the
magnetic neutral line from Wilcox Solar Observatory synoptic maps.\footnote{\url{http://wso.stanford.edu}}
Low-frequency LOFAR monitoring showed that the amplitude of the
$r^{-2}$ model must be allowed to vary in time \citep{Tiburzi2021}.
Gaussian-process descriptions of this time-variable amplitude have
since been developed \citep{Hazboun2022,Nitu2024,Susarla2024}. Applied
to eleven years of LOFAR data, they show that pulsars within
$3^\circ$ of the ecliptic see systematically higher densities than
those farther away \citep{Susarla2024}. The Crab pulsar, at ecliptic
latitude $\beta\approx-1.3^\circ$, lies in this regime, and spherical
models are known to underfit such lines of sight \citep{Susarla2025}.
For Faraday rotation, simultaneous DM and RM measurements of
millisecond pulsars have constrained the solar-wind magnetic field at
6--10 solar radii \citep{You2012}. \citet{Ord2007} measured an RM
change of $\sim160~\mathrm{rad\,m^{-2}}$ for PSR J1801$-$2304 at
$0.95^\circ$ elongation and compared it with source-surface models.
\citet{Kooi2022} review current coronal and solar-wind Faraday-rotation
techniques.

Radio-wave propagation through turbulent ionized plasma can produce
phase fluctuations, angular broadening, scintillation, and multipath
propagation, providing a probe of electron-density fluctuations along
the line of sight \citep{Rickett1990}. Radio scattering and
scintillation have also been used to investigate density fluctuations
and turbulence in the inner solar wind \citep{HarmonColes2005JGR}.
Low-frequency (30--80~MHz) imaging of the Crab Nebula during its 2024
conjunction showed angular broadening elongated perpendicular to the
radial direction, becoming more anisotropic closer to the Sun, and
substructure in the broadened image when the line of sight crossed a
coronal streamer \citep{Zhang2025ApJ}.

Long-term monitoring also demonstrates that propagation through the
Crab Nebula contributes to the observed variability.
\citet{McKee2018} measured correlated changes in scattering and DM,
separating scattering delays from dispersive delays in their timing
analysis. \citet{Driessen2019} observed evolving echoes and changes
in scatter broadening, including both exponential and non-exponential
echo profiles. These results establish a variable nebular propagation
background that must be considered when interpreting conjunction
observations. More recent CHIME monitoring reveals a diverse population
of plasma-lensing echoes in the same observing band used here
\citep{SerafinNadeau2026}.

The observations considered here differ from previous Crab
solar-conjunction measurements in both observing frequency and
strategy. We use five years of 400--800~MHz DM and RM measurements
from CHIME. Our goals are to characterize the DM and RM variations near
conjunction, to estimate the line-of-sight magnetic field from them,
and to examine the secondary DM enhancement of 2025.


\section{Solar Conjunction Geometry}
\label{subsec:conjunction}

We define the annual conjunction time as the UTC epoch at which the
geocentric angular separation between the Sun and the Crab pulsar
reaches its annual minimum. In each year from 2022
through 2026 this minimum, $1.29^\circ$, occurs on 15 June (UTC),
between 00:51 (2024) and 18:40 (2023). CHIME observes the Crab once a
day, at transit, which near conjunction coincides with local solar noon
at the observatory (about 19:55~UTC), so the observation closest to the Sun
is on 15 June in 2022, 2023, and 2026 (elongation
$1.29^\circ$--$1.33^\circ$) but on 14 June in 2024 ($1.31^\circ$,
compared with $1.50^\circ$ on 15 June); in 2025 the observations of
14 and 15 June are at almost the same elongation ($1.37^\circ$ and
$1.39^\circ$). We use 15 June as the reference date for all five
conjunctions, and call 10--20 June, when the elongation at transit is
below about $5.7^\circ$, the conjunction window.


\section{Dispersion Measure}
\label{sec:dm}


\subsection{Method}
\label{subsec:dm_method}

We analyze Crab pulsar DM measurements spanning the annual solar
conjunctions from 2022 through 2026. The historical CHIME monitoring
archive is described by \citet{SerafinNadeau2026}, whose published
daily stacks span late October 2021 through March 2024. Later
observations, including those of June 2024, 2025, and 2026, were
processed with the same pipeline; daily stacks were not produced for
them, and they have not been published separately.

\paragraph{Data.}
For each daily transit we use baseband single-pulse data recorded by
CHIME and processed by the CHIME Crab pulsar monitoring pipeline
\citep{SerafinNadeau2026}. These products are used for the DM analysis
in all five years and for the RM analysis in 2026. Each observation
contains up to 512 triggered pulses, each a
$\sim38$~ms segment of channelized complex voltages in two linear
polarizations, with 1024 frequency channels spanning 400--800~MHz and
a time resolution of 2.56~$\mu$s. Each pulse segment has already been
dedispersed by the processing pipeline at a reference value
$\DM_{\mathrm{ref}}$, the DM that maximizes the peak of the
frequency-averaged pulse profile for that observation, determined after
the bandpass has been flattened using the background emission of the
Crab Nebula. The voltages are not polarization-calibrated. We measure the
residual dispersion $\delta\DM$ relative to this reference, and take
$\DM_{\mathrm{obs}}=\DM_{\mathrm{ref}}+\delta\DM$. Because the pulses
are dedispersed at $\DM_{\mathrm{ref}}$, any error in the reference
value appears directly in $\delta\DM$ and cancels in
$\DM_{\mathrm{obs}}$. We use only main-pulse triggers, as classified
in the pipeline pulse table.

\paragraph{Subband profiles.}
For each pulse we form the total intensity $I=|X|^2+|Y|^2$ and
normalize each frequency channel by the median and robust standard
deviation ($1.4826\times$ the median absolute deviation) of its
off-pulse samples. Channels with zero or non-finite off-pulse
variance are discarded; these are the channels flagged upstream,
which carry constant values. The remaining channels are averaged into
eight subbands of 50~MHz. For each pulse we measure the
signal-to-noise ratio (SNR) of the band-summed profile after smoothing over
eight samples. Only pulses with ${\rm SNR}\geq8$ are retained, because
aligning weak pulses to a template locks onto noise and biases the
differential delays toward zero. The retained pulses are aligned to
each other with a single integer-sample shift per pulse, which does
not change the delays between subbands, and are then stacked.

\paragraph{Joint dispersion--scattering fit.}
The Crab pulsar shows strong and variable scattering in this band, and
the scattering tail shifts the centroid of the low-frequency profiles.
Cross-correlating subband profiles therefore mixes scattering into the
inferred dispersion. To separate the two, we fit all subband
profiles of the stacked pulse simultaneously with an exponentially
modified Gaussian, i.e.\ a Gaussian of width $\sigma_i$ convolved with
a one-sided exponential of timescale $\tau_i$:
\begin{equation}
    P_i(t)=A_i\,\left[\mathcal{G}(t-\mu_i;\sigma_i)\ast
    \frac{e^{-t/\tau_i}}{\tau_i}\right]+b_i ,
    \label{eq:emg}
\end{equation}
with
\begin{align}
    \mu_i &= t_{\mathrm{ref}} + k_{\mathrm{DM}}\,\delta\DM\,
             \left(\langle\nu^{-2}\rangle_i-\overline{\langle\nu^{-2}\rangle}\right),
    \label{eq:dm_delay}\\
    \tau_i &= \tau_{600}\left(\nu_i/600~\mathrm{MHz}\right)^{-\alpha},
    \label{eq:tau_scaling}\\
    \sigma_i^2 &= \sigma_0^2+\frac{1}{12}\left[k_{\mathrm{DM}}\,\delta\DM\,
             \left(\nu_{i,\mathrm{lo}}^{-2}-\nu_{i,\mathrm{hi}}^{-2}\right)\right]^2 ,
    \label{eq:smear}
\end{align}
where $k_{\mathrm{DM}}=4.148808\times10^{3}~\mathrm{s\,MHz^{2}\,pc^{-1}\,cm^{3}}$,
$\langle\nu^{-2}\rangle_i$ is the mean of $\nu^{-2}$ over the retained
channels of subband $i$, $\overline{\langle\nu^{-2}\rangle}$ is the mean
of $\nu^{-2}$ over all retained channels, $\nu_i$ is the center frequency
of subband $i$, and $\nu_{i,\mathrm{lo}}$ and $\nu_{i,\mathrm{hi}}$ are its
lower and upper edges. The last term accounts for dispersive smearing
within each subband. The free parameters are $t_{\mathrm{ref}}$,
$\delta\DM$, $\sigma_0$, $\tau_{600}$, and a per-subband amplitude
$A_i$ and baseline $b_i$. We fix $\alpha=4$, the value for a thin screen with a square-law phase
structure function,
which is consistent with the index $\alpha=3.9\pm0.5$ measured for the
Crab by \citet{Driessen2019}. If $\alpha$ is left free it can fall toward
$\alpha\approx2$, where the scattering delay becomes degenerate with
the $\nu^{-2}$ dispersive delay. The fit covers $-256$ to
$+640~\mu$s around the pulse, so that the scattering tail is included.
Residuals are weighted by the off-pulse noise of each subband, and we
use a soft-$\ell_1$ robust loss to limit the influence of narrowband
interference.

\paragraph{Uncertainties and validation.}
The statistical uncertainty on $\delta\DM$ is the half-width of the
16th--84th percentile interval from 300 bootstrap resamplings of the
pulses. As a consistency check we fit the even- and odd-numbered pulses
separately. An observation is discarded if the two halves differ by
more than five times the bootstrap uncertainty, if its uncertainty
exceeds five times the median for that year, or if $|\delta\DM|$
approaches the fit bound of $0.03~\mathrm{pc\,cm^{-3}}$.

We tested the fit on simulated observations processed with the same
code (Appendix~\ref{appendix:validation}). For
$\tau_{600}=10$--$70~\mu$s, which covers the 25--67~$\mu$s observed in
2025 (Section~\ref{subsec:dm_results}), it recovers injected values of
$\delta\DM=0$--$7\times10^{-3}~\mathrm{pc\,cm^{-3}}$ to within
$0.06\times10^{-3}~\mathrm{pc\,cm^{-3}}$, and the bias does not change
with $\tau_{600}$. Profile cross-correlation instead overestimates
$\delta\DM$ by $(0.9$--$1.6)\times10^{-3}~\mathrm{pc\,cm^{-3}}$, by
more for stronger scattering. Because $\Delta\DM$ is measured against
a baseline from days with $\tau_{600}\approx25$--$30~\mu$s, what matters
is how the bias changes with $\tau_{600}$. Between 30 and $70~\mu$s it
changes by less than $0.01\times10^{-3}~\mathrm{pc\,cm^{-3}}$ when
$\alpha=4$ is correct. It changes by at most
$0.05\times10^{-3}~\mathrm{pc\,cm^{-3}}$ if the true index is 3.5 or 4.4,
or if the tail has a second, longer component. This is about a
quarter of the smallest yearly $\sigma_{\Delta\DM}$.

\paragraph{Background subtraction.}
The observed DM includes contributions from the interstellar medium,
the Crab Nebula, and the solar wind. To characterize the excess near
conjunction relative to a slowly varying background, for each year we
fit an unweighted straight line to $\DM_{\mathrm{obs}}$ on the days
outside 10--20 June (the baseline days),
\begin{equation}
    \DM_{\mathrm{bg}}(t)=at+b ,
    \label{eq:dm_background}
\end{equation}
and define the background-subtracted DM as
\begin{equation}
    \Delta\DM(t)=\DM_{\mathrm{obs}}(t)-\DM_{\mathrm{bg}}(t).
    \label{eq:delta_dm}
\end{equation}
Solar-wind structure is not confined to 10--20 June. A day outside
the window that carries a real enhancement would pull the line and,
through Equation~(\ref{eq:dm_sys}) below, inflate the uncertainties
against which it is itself judged. We therefore remove baseline days
one at a time while the largest residual exceeds four times the robust
scatter, refitting after
each removal. This removes 21 and 22 June in 2025 and no day in
2022--2024. In 2026 the excess after conjunction decays slowly and is
still $(2.0$--$3.7)\times10^{-3}~\mathrm{pc\,cm^{-3}}$ on 21--24 June.
A gradual tail like this does not
trip the outlier criterion, so we exclude 21--24 June 2026 by hand.
Days removed in this way are still measured and shown, but they are
not used for the baseline, for $\sigma_{\mathrm{sys}}$, or for the null
test of Section~\ref{subsec:dmrm_method}.
This subtraction removes the fitted secular trend, but it does not
separate solar-wind variations from short-term nebular variations.
The scatter of the baseline days about the line exceeds the
bootstrap uncertainties, by a reduced $\chi^2$ of order $10^2$--$10^3$.
This excess reflects day-to-day DM and profile variations of the
pulsar and nebula, which the per-observation bootstrap does not
capture. We therefore add in quadrature an excess term
\begin{equation}
    \sigma_{\mathrm{sys}}^2=\max\left(0,\;s_{\mathrm{off}}^2-
    \mathrm{median}\left[\sigma_{\mathrm{stat}}^2\right]_{\mathrm{off}}\right),
    \label{eq:dm_sys}
\end{equation}
where $s_{\mathrm{off}}$ is the standard deviation of the baseline-day
residuals. The resulting uncertainties are
$\sigma_{\Delta\DM}\approx(0.19$--$0.69)\times10^{-3}~\mathrm{pc\,cm^{-3}}$,
depending on the year (0.29, 0.69, 0.43, 0.19, and 0.60 for 2022--2026).

\begin{figure}[ht!]
    \centering
    \includegraphics[width=\columnwidth]{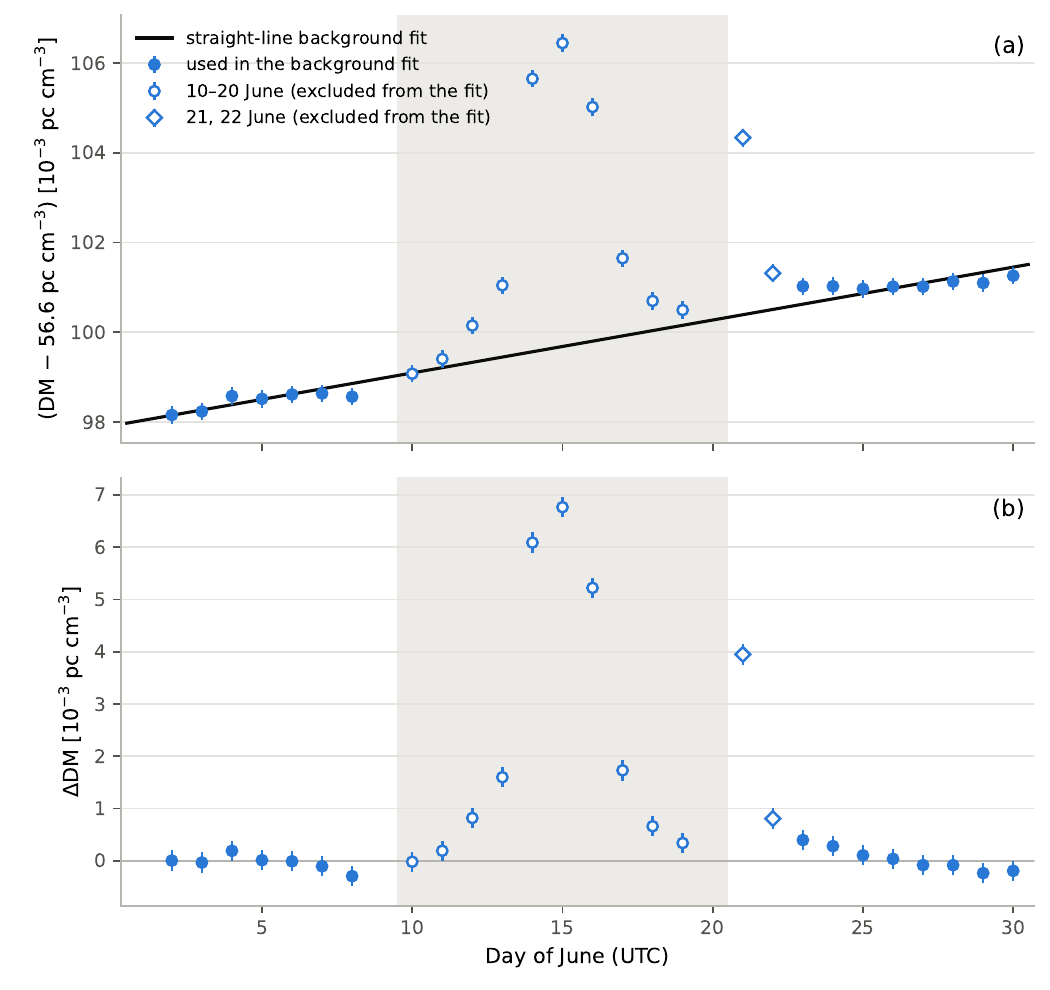}
    \caption{DM of the Crab pulsar around the 2025 conjunction.
    (a) Observed DM before background subtraction. The line is the
    straight-line background fit to the filled points. Open circles mark
    10--20 June (shaded), and open diamonds mark 21--22 June, which are
    excluded from the fit. (b) Background-subtracted DM, $\Delta\DM$,
    with the same symbols as in (a).}
    \label{fig:dm2025}
\end{figure}

Figure~\ref{fig:dm2025}(a) shows the observed 2025 DM before
background subtraction, together with the fitted background line. The
DM rises toward conjunction and returns to the background afterwards.
Figure~\ref{fig:dm2025}(b) shows the background-subtracted DM for
2025.

\subsection{Multi-Year DM Variations}
\label{subsec:dm_results}

The DM time series from 2022 through 2026 show enhancements near the
annual Crab pulsar conjunction.

\begin{figure*}[ht!]
    \centering
    \includegraphics[width=0.85\textwidth]
    {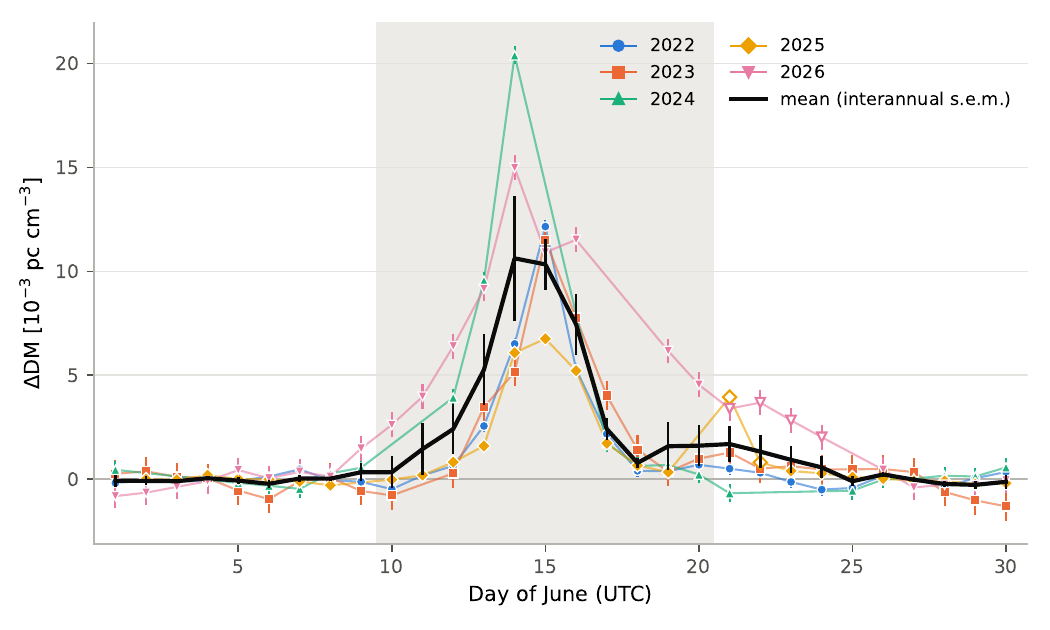}
    \caption{Background-subtracted DM around the annual Crab pulsar
    conjunctions from 2022 through 2026. Because the minimum solar
    elongation falls on 15 June in every year, the five years are plotted
    on a common day-of-June axis. The black curve is the mean over the
    years available on each day, with the interannual standard error. The
    shaded interval, 10--20 June, is excluded from the background fits.
    Open symbols mark days outside the window that are also excluded
    (21--22 June 2025 and 21--24 June 2026). Days that fail the quality
    criteria of Section~\ref{subsec:dm_method} are not shown.
    }
    \label{fig:dm_all_years}
\end{figure*}

The DM peak in each year (Figure~\ref{fig:dm_all_years}) occurs near
the expected annual conjunction time; it falls on 14 June in 2024 and 2026 and on
15 June in the other years. 2025 is the only year in which a separate
enhancement appears after $\Delta\DM$ has returned to the background:
$\Delta\DM$ falls to $(0.34\pm0.19)\times10^{-3}~\mathrm{pc\,cm^{-3}}$ on
19 June and rises again two days later. In 2026, by contrast, the
excess decays slowly after conjunction and is still
$(2.0$--$3.7)\times10^{-3}~\mathrm{pc\,cm^{-3}}$ on 21--24 June. The first and stronger peak,
$\Delta\DM=(6.76\pm0.20)\times10^{-3}~\mathrm{pc\,cm^{-3}}$ on 15 June,
occurs at the expected conjunction date. The second occurs on
21 June 2025, with
$\Delta\DM=(3.95\pm0.20)\times10^{-3}~\mathrm{pc\,cm^{-3}}$, the largest
value outside the conjunction window. A smaller excess remains on 22 June,
$(0.81\pm0.20)\times10^{-3}~\mathrm{pc\,cm^{-3}}$, and 23 June is at the
$2\sigma$ level. The DM on 20 June fails the criteria of
Section~\ref{subsec:dm_method}, so the rise is not sampled.
The quoted uncertainties include the excess scatter of the baseline days
(Equation~\ref{eq:dm_sys}), from which 21 and 22 June are excluded. The
21 June value is about 20 times that uncertainty, but temporal
correlations in the residuals mean that this ratio should not be read
as a formal detection significance.

The 2025 conjunction peak is also low compared with those of the
other four years. The 2025 peak of
$6.8\times10^{-3}~\mathrm{pc\,cm^{-3}}$ compares with peak values of
12.2 (2022), 11.5 (2023), 20.4 (2024), and 15.0 (2026)
$\times10^{-3}~\mathrm{pc\,cm^{-3}}$.
The baseline days are not free of solar wind. For an $n_e\propto r^{-2}$
wind the column along the line of sight scales as
$(\pi-\epsilon)/\sin\epsilon$ at elongation $\epsilon$, so the baseline
days, at $\epsilon\approx5.4^\circ$--$15^\circ$, carry 8--23\% of the
column at minimum elongation. The straight-line fit absorbs part of this,
and $\Delta\DM$ therefore underestimates the solar-wind column by a
similar fraction. Because the observing geometry repeats from year to
year, this affects all five years in nearly the same proportion and does
not by itself explain the low 2025 peak. The mean conjunction DM varies
from year to year, although no clear long-term trend is apparent from
the five annual conjunctions currently available.

\paragraph{Interpretation.}

The increase in DM toward conjunction is consistent with the
expectation that the electron column density along the Crab line of
sight increases as the solar elongation decreases
\citep{You2012,Tokumaru2022SoPh}. A low peak in a single year does not
by itself require unusual structure. In 2024, the year with the
largest peak, low-frequency images of the Crab Nebula show that the
line of sight on 13 June passed through a more pronounced streamer than
on 15 June \citep{Zhang2025ApJ}, which may contribute to the high
2024 values. LOFAR studies find that the
density normalization of the solar wind varies from year to year,
with the largest scatter for lines of sight close to the ecliptic
\citep{Tiburzi2021,Susarla2024}. A fixed $r^{-2}$ profile would
therefore not be expected to predict the relative peak heights in
Figure~\ref{fig:dm_all_years}.

Two features of the 2025 conjunction stand out: its low peak
amplitude and its secondary enhancement. Because both could have arisen from
the peak-maximizing estimator used for $\DM_{\mathrm{ref}}$, we
compare that estimator with our joint dispersion--scattering fit. In
2025 the residual correction $\delta\DM$ lies between $-2.4$ and
$+0.4\times10^{-3}~\mathrm{pc\,cm^{-3}}$ on every day, so both
features are present in both estimates. The joint fit does, however,
give a smaller secondary enhancement than the peak-maximizing
estimator, and one that is more concentrated on 21 June. By contrast, the same comparison
identifies a pipeline error of $\approx0.012~\mathrm{pc\,cm^{-3}}$ on
11 June 2023, which the joint fit corrects. Scattering was strong
throughout June 2025. The fitted $\tau_{600}$ is 25--30~$\mu$s on
off-window days and rises to 33--67~$\mu$s near conjunction, with the
largest value on 19 June. Our profile model does not fully describe
the observed pulse shapes at this SNR, giving reduced
$\chi^2$ of 17--56 per observation. The DM errors from this mismatch
are absorbed by the excess off-window term, but they are not modeled
explicitly.

An additional complication is propagation through the Crab Nebula.
CHIME monitoring has revealed plasma-lensing echoes produced by
nebular material \citep{SerafinNadeau2026}. Long-term observations
also show correlated scattering and DM changes attributable to nebular
structures \citep{McKee2018}, as well as evolving echo morphologies
\citep{Driessen2019}. Non-exponential echoes are not captured by the
model of Equation~(\ref{eq:emg}). A linear background subtraction
therefore does not demonstrate that every residual feature originates
in the solar wind.

The origin of the secondary enhancement is not determined by the
current analysis. Additional years of DM measurements will be useful
for determining how frequently similar structures occur.


\section{Rotation Measure and the Line-of-Sight Magnetic Field}
\label{sec:rm}


\subsection{Method}
\label{subsec:rm_method}

\paragraph{Stokes spectra.}
For 2026 we use the same single-pulse products as for the DM analysis.
For 2022--2025 we use channelized single-pulse voltages in the same
format (prepared by H.S.). In both cases the voltages are not
polarization-calibrated: no correction has been applied for the
relative gain, cross-hand phase, or leakage of the two feeds. For each
pulse we form
$I=|X|^2+|Y|^2$, $Q=|X|^2-|Y|^2$, $U=2\,\mathrm{Re}(XY^{*})$, and
$V=2\,\mathrm{Im}(XY^{*})$ from the two linear-feed voltages. The
on-pulse window is the contiguous region around the peak of the
channel-median intensity time series that lies above half of its
maximum. The off-pulse region is everything outside a guard interval
around that window. $Q$, $U$ and $V$ are on-pulse means with their
off-pulse means subtracted, normalized by
$\sqrt{\langle|X|^2\rangle_{\mathrm{off}}\langle|Y|^2\rangle_{\mathrm{off}}}$.
Channels with non-finite values or zero off-pulse variance are excluded,
as are channels that are valid in fewer than 90\% of the pulses in an
observation. Main-pulse and interpulse triggers are fitted in the same
way and combined; most of the pulses (about 85--90\%) are main-pulse
triggers.

\paragraph{Rotation and cross-hand-delay model.}
For trial values of RM and of an instrumental cross-hand delay
$\tau_{\mathrm{xy}}$, each channel's polarization vector is first
rotated in the $U$--$V$ plane by
$\phi_{\mathrm{xy}}=2\pi\tau_{\mathrm{xy}}(\nu-\bar\nu)$, and then in
the $Q$--$U$ plane by $2\RM(\lambda^2-\overline{\lambda^2})$, where
$\bar\nu$ and $\overline{\lambda^2}$ are the means of $\nu$ and
$\lambda^2$ over the retained channels. The
figure of merit is the coherence of the derotated, unit-normalized
polarization vectors $\hat{\mathbf{p}}(\nu)$,
\begin{equation}
    \mathcal{L}(\RM,\tau_{\mathrm{xy}})=2-2\left|\big\langle
    \hat{\mathbf{p}}(\nu)\big\rangle_{\nu}\right| ,
    \label{eq:rm_loss}
\end{equation}
which is minimized when all channels share a common polarization
direction. Normalizing each channel removes the dependence on spectral
amplitude.

\paragraph{Combining pulses.}
Individual pulses have too little polarized signal for a reliable RM;
single-pulse fits frequently jump between competing minima, among them
a solution near zero RM discussed below. For every pulse we evaluate
$\mathcal{L}$ on a grid with $|\RM|\le100~\mathrm{rad\,m^{-2}}$
(step 1~rad\,m$^{-2}$) and $|\tau_{\mathrm{xy}}|\le3$~ns (step 0.1~ns)
and refine the minimum with a parabolic interpolation, which gives a
single-pulse $\RM_i$, $\tau_{\mathrm{xy},i}$ and minimum loss
$\mathcal{L}_i$. Minima with $|\RM_i|\le5~\mathrm{rad\,m^{-2}}$ are
discarded at this stage. After the pulse selection described next,
the RM of an observation is the centre of the dominant peak of the
retained $\RM_i$. We find the 12~rad\,m$^{-2}$-wide interval that
contains the most retained pulses and take the median of the pulses in
it; this median defines the centre of a new 12~rad\,m$^{-2}$ window,
whose median is taken in turn, and the step is repeated until it
converges. Using the peak rather than the mean of all pulses makes the
estimate insensitive to the minority of pulses whose fit has jumped to
another minimum.

\paragraph{Rejecting leakage-dominated pulses.}
We classify each pulse by its fitted delay and coherence. Outside the
conjunction window the single-pulse minima fall into two populations
that are well separated in cross-hand delay: pulses at the baseline RM
have $|\tau_{\mathrm{xy}}|$ with yearly medians of 0.75--0.90~ns
(robust scatter 0.2--0.34~ns), while pulses in the near-zero solution
discussed under the quality criteria below have
$|\tau_{\mathrm{xy}}|=0.24$--0.46~ns. We keep a pulse only if its
fitted delay lies within 0.3~ns of the year's median for the
baseline-RM population and closer to it than to the near-zero
population, and if $\mathcal{L}_i<1.5$, i.e.\ if it shows coherent
polarization. The selection uses only the delay and the coherence, not
the fitted RM, so it does not pull the retained RM toward the baseline.
On a typical day it keeps about half of the pulses. Near conjunction
it keeps fewer, about 15--20\% within $7~R_\odot$ of the Sun: the
increased system temperature lowers the polarized SNR, so more pulses
fail the coherence cut, and the delays of the remaining ones scatter
more. The selection also removes intervals within a single transit,
tens of pulses long, in which the near-zero solution dominates;
without it these intervals produce secondary peaks in the single-pulse
RM distribution at a nearly fixed $\RM\approx-8$~rad\,m$^{-2}$,
independent of the day's true RM.

\paragraph{Uncertainties and quality criteria.}
The statistical uncertainty is the 16th--84th percentile half-width
from 300 bootstrap resamplings of the retained pulses. We discard an
observation if any of the following holds:
(i) fewer than 90\% of bootstrap solutions lie within half an RMSF
width of the best fit, where
$\mathrm{FWHM}_{\mathrm{RMSF}}=2\sqrt3/\Delta\lambda^2\approx8.4~\mathrm{rad\,m^{-2}}$
for the $\lambda^2$ range of the channels used (8.2~rad\,m$^{-2}$ for the
full 400--800~MHz band);
(ii) fewer than eight pulses are retained, or fewer than 40\% of them
lie within the 12~rad\,m$^{-2}$ window about
the converged median; or
(iii) the even- and odd-numbered pulses give RMs that differ by more
than $\mathrm{FWHM}_{\mathrm{RMSF}}/4$.
The near-zero solution has no $\lambda^2$ rotation, and it comes with
an anomalous cross-hand delay. Because $Q$, $U$, and $V$ are on-pulse
minus off-pulse quantities, steady emission, including that of the
Sun, subtracts out; only contamination that is itself pulsed survives.
We attribute the component to leakage of the pulsar's total intensity
into $Q$ and $U$ in the uncalibrated feeds, which is not rotated by
$\lambda^2$. Near the Sun the increased system temperature lowers the
SNR of the true polarized signal and can therefore make this component
dominant, but the Sun is not its source. Within one RMSF width of zero
a true RM cannot be separated from this component by its RM alone. We
therefore show as an open symbol, and exclude from the fits of
Section~\ref{subsec:dmrm_method}, any observation whose RM before
ionospheric correction lies within $\mathrm{FWHM}_{\mathrm{RMSF}}$ of
zero, or within 15~rad\,m$^{-2}$ of zero while the pulses in its
dominant peak, before the delay selection, have a median delay closer
to that of the near-zero solution than to the year's median. Together
with the $|\RM_i|\le5~\mathrm{rad\,m^{-2}}$ cut, this means that
genuine astrophysical values near zero RM before ionospheric
correction cannot be recovered.

\paragraph{Ionospheric correction.}
We compute the ionospheric RM toward the Crab from CHIME with
\texttt{spinifex},\footnote{\url{https://git.astron.nl/RD/spinifex}}
using the UQRG global ionospheric maps \citep{Orus2005UQRG} (a
single-layer model at 450~km) and the IGRF geomagnetic field
\citep{Alken2021IGRF}. The ionospheric RM is
evaluated every 2~minutes across each observation and averaged. Its
value is 1.7 to 5.6~rad\,m$^{-2}$, with a model uncertainty of
$\simeq0.13~\mathrm{rad\,m^{-2}}$ that rises to $\simeq0.94~\mathrm{rad\,m^{-2}}$
on two days in 2022. Before correction, the off-window RM residuals correlate with the
predicted ionospheric RM (correlation coefficients 0.77--0.90 in the
five years). The correction reduces their scatter from 0.27--0.72 to
0.16--0.34~rad\,m$^{-2}$, and the off-window reduced $\chi^2$ computed
with the statistical uncertainties alone from 8--35 to 2--8.
We note that the measured RM grows with the ionospheric RM; a sign
error in the polarization convention would reverse this correlation.

\paragraph{Background subtraction and error budget.}
For each year, the baseline is the median of the
ionosphere-corrected RM over usable days outside 10--20 June, and
$\Delta\RM$ is measured relative to it. The uncertainty on each daily
$\Delta\RM$ is
\begin{equation}
    \sigma_{\Delta\RM}^2=\sigma_{\mathrm{stat}}^2+\sigma_{\mathrm{iono}}^2
    +\sigma_{\mathrm{sys}}^2 ,
    \label{eq:rm_err}
\end{equation}
where $\sigma_{\mathrm{sys}}$ is defined as in Equation~(\ref{eq:dm_sys}),
but with the mean rather than the median of
$\sigma_{\mathrm{stat}}^2+\sigma_{\mathrm{iono}}^2$ over the off-window
days. Typical values are $\sigma_{\Delta\RM}\approx0.2$--0.35~rad\,m$^{-2}$,
rising to $\approx1$~rad\,m$^{-2}$ on the two days in 2022 with large
ionospheric uncertainty. The off-window baselines change from $-46.7$
(2022) to $-40.2~\mathrm{rad\,m^{-2}}$ (2025--2026). The median
single-pulse cross-hand delay changes by $\lesssim0.15$~ns over the same
period,
which is too little to produce a 6~rad\,m$^{-2}$ shift, so
the change in baseline is plausibly nebular in origin. It does not
affect $\Delta\RM$.


\subsection{Line-of-Sight Field from Joint DM and RM Variations}
\label{subsec:dmrm_method}

Combining Equations~(\ref{eq:dm}) and~(\ref{eq:rm}), the ratio of the
solar-wind contributions gives the electron-density-weighted mean
line-of-sight field,
\begin{equation}
    \langle B_\parallel\rangle=\frac{\int n_eB_\parallel\,dl}{\int n_e\,dl}
    =1.232~\mu\mathrm{G}\;\frac{\Delta\RM/(\mathrm{rad\,m^{-2}})}{\Delta\DM/(\mathrm{pc\,cm^{-3}})} .
    \label{eq:bpar}
\end{equation}
We estimate this ratio as the slope of $\Delta\RM$ against $\Delta\DM$.
Both quantities have comparable fractional uncertainties, so we fit
with orthogonal distance regression, which accounts for errors on both
axes; ordinary least squares in $\Delta\RM$ would bias the slope low.
Because both quantities are background subtracted, the line is
constrained to pass through the origin. This assumes that the
solar-wind contributions left in the two baselines, a straight line
for DM (Section~\ref{subsec:dm_method}) and a median for RM, are in
the same proportion as the excesses; a residual solar-wind column in
one baseline that is not matched in the other would bias the slope. As a check, a fit with a free
intercept returns intercepts consistent with zero in all but one sector
(Section~\ref{subsec:dmrm_results}). The quoted slope
uncertainty is the larger of the regression uncertainty and the
bootstrap uncertainty from resampling days.

Days in 10--20 June are fitted as one group, unless $\Delta\RM$
changes sign within that window. A sign change indicates that the line
of sight has crossed a magnetic-sector boundary, and a single line
would average the opposite polarities toward zero. In those years we split the window at the sign changes: at 13.5 June
in 2023, at 12.5 and 16.5 June in 2024, and at 13.5, 14.5, 15.5 and
17.5 June in 2026. Where only one day of a sector has usable DM, we
quote the ratio $\Delta\RM/\Delta\DM$ for that day instead of a fit. Days outside the window, except those
removed from the DM baseline (Section~\ref{subsec:dm_method}), serve as
a null test; their slope should be consistent with zero. Equation~(\ref{eq:bpar})
underestimates the local field strength wherever $B_\parallel$ reverses
along the path. Because the weighting is by $n_e$, the result is
dominated by the few days of largest $\Delta\DM$, i.e.\ the closest
approach to the Sun. Those same days are the ones most often excluded by the RM criteria
of Section~\ref{subsec:rm_method}, although the per-pulse delay
selection recovers several of them.


\subsection{Results}
\label{subsec:dmrm_results}

\begin{figure*}[ht!]
    \centering
    \includegraphics[width=0.95\textwidth]{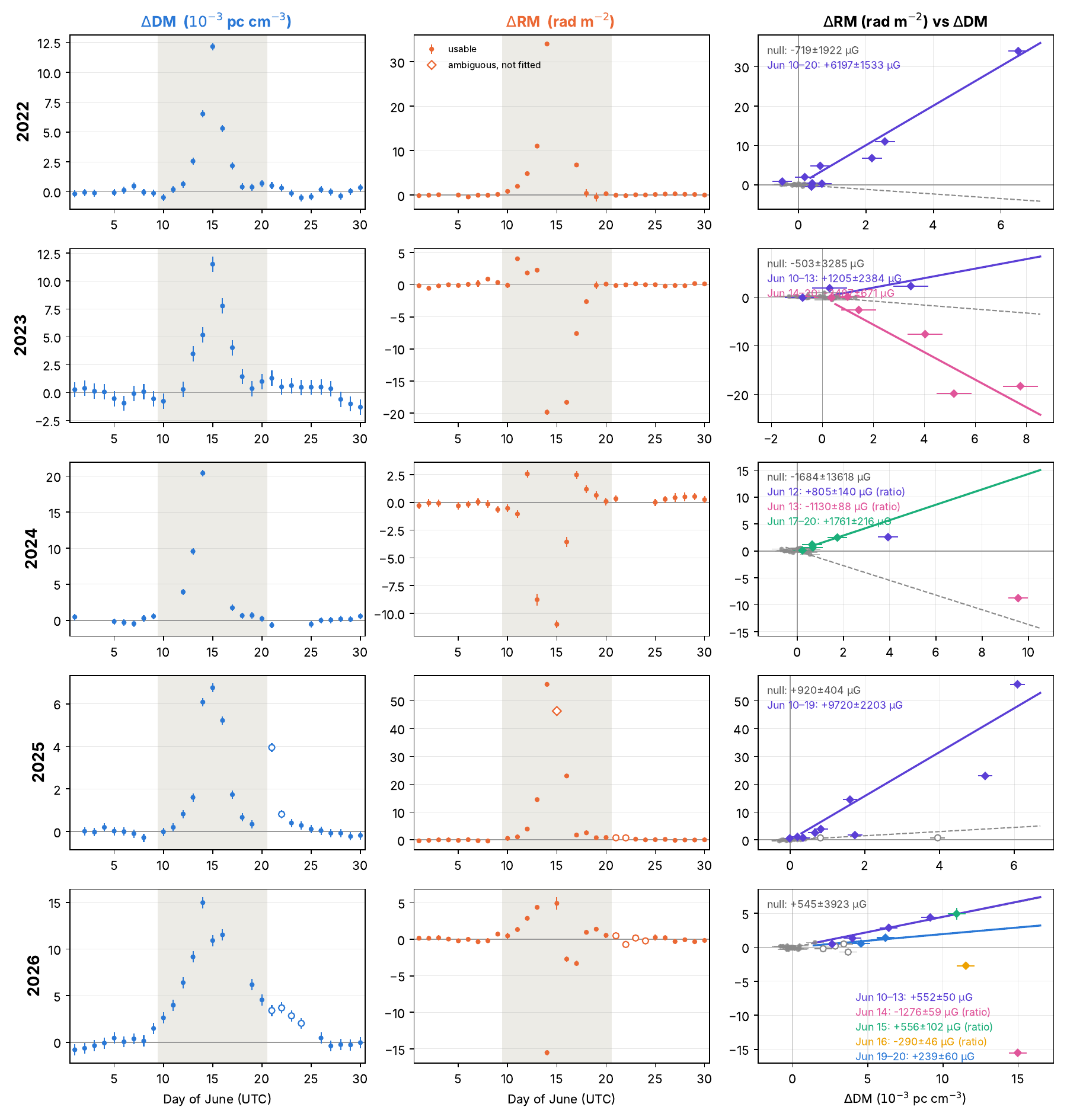}
    \caption{The Crab pulsar during the June solar conjunctions of
    2022--2026, one row per year. Left: background-subtracted DM from the
    joint dispersion--scattering fit (Section~\ref{subsec:dm_method}).
    Middle: ionosphere-corrected, background-subtracted RM from the
delay-selected single pulses (Section~\ref{subsec:rm_method}). The open
diamond (15 June 2025) is consistent with contamination by the
near-zero-RM solution and is not used in the fits. The shaded interval,
    10--20 June, is excluded from both background fits. Right: $\Delta\RM$
    versus $\Delta\DM$, with orthogonal-distance-regression fits through
    the origin for each magnetic sector (colors; single-day sectors are ratios and are not drawn as lines) and for the
    out-of-window null test (gray, dashed); the labels give the
    corresponding $\langle B_\parallel\rangle$ from
    Equation~(\ref{eq:bpar}). Open symbols mark 21--22 June 2025 and
    21--24 June 2026, which are excluded from the background fits and
    from the null test.  Error bars include the statistical,
    ionospheric, and excess baseline-day terms. Days that fail the quality
    criteria are not shown.}
    \label{fig:year_panels}
\end{figure*}

Figure~\ref{fig:year_panels} shows $\Delta\DM$, $\Delta\RM$, and their
correlation for each conjunction. After the ionospheric correction
(Section~\ref{subsec:rm_method}), the off-window reduced $\chi^2$ of
$\Delta\RM$ is 0.7--1.0 in every year.

Every year shows an RM excursion near conjunction. The largest
excursions are $+34.0$~rad\,m$^{-2}$ on 14 June 2022,
$-19.8$~rad\,m$^{-2}$ on 14 June 2023, $-11.0$~rad\,m$^{-2}$ on 15 June
2024, $+55.9$~rad\,m$^{-2}$ on 14 June 2025, and $-15.5$~rad\,m$^{-2}$
on 14 June 2026. In 2023, 2024, and 2026, $\Delta\RM$ changes sign
within the conjunction window. In 2024 $\Delta\RM$ is positive on
12 June ($+2.6$~rad\,m$^{-2}$), negative on 13, 15 and 16 June ($-8.8$,
$-11.0$ and $-3.6$~rad\,m$^{-2}$), and positive again from 17 June. The
DM on 15 and 16 June fails the criteria of
Section~\ref{subsec:dm_method}, so the negative sector enters the fit
only through 13 June. In 2026 $\Delta\RM$ is positive through 13 June,
$-15.5$~rad\,m$^{-2}$ on 14 June, $+4.9$~rad\,m$^{-2}$ on 15 June,
negative on 16--17 June ($-2.7$ and $-3.3$~rad\,m$^{-2}$), and positive
again from 18 June. The DM on 17 and 18 June fails the criteria of
Section~\ref{subsec:dm_method}, so the 14, 15 and 16 June sectors are
single days and only 19--20 June of the last sector enters the fit
(Table~\ref{tab:bpar}).

The per-pulse delay selection (Section~\ref{subsec:rm_method})
recovers four days near closest approach that would otherwise have no
usable RM: 14 June 2022 ($\Delta\RM=+34.0$~rad\,m$^{-2}$), 16 June 2023
($-18.3$, close to the value on 14 June), 13 June 2024 ($-8.8$, from
16 retained pulses), and 14 June 2026 ($-15.5$). On 14 June 2022 the
retained pulses have the normal cross-hand delay (0.82~ns) and the RM
is constant across the transit. Its RM before ionospheric correction,
$\approx-9.8$~rad\,m$^{-2}$, lies just outside one RMSF width of zero
and close to the secondary peak produced by the near-zero solution, so
the delay is what distinguishes it. On 14 June 2026 the RM is constant
over the first part of the transit, which contains most of the
retained pulses, and scattered in the remainder, where few pulses
survive. On 15 June 2025 $\Delta\RM\approx+46$~rad\,m$^{-2}$, between
the values on 14 and 16 June. Its RM before correction
($\approx+8.8$~rad\,m$^{-2}$) is at the edge of the RMSF width and its
delay is intermediate (0.54~ns), so we show it as an open symbol and do
not fit it. The days still without usable RM are 15--16 June 2022,
15 June 2023 and 14 June 2024, on each of which fewer than eight pulses
survive the selection. These are not always the day of minimum
elongation; the RM on 15 June is usable in 2024 and 2026.

Unlike DM, the 2025 RM series shows no secondary feature. On 21 June
2025, when $\Delta\DM$ reaches its secondary maximum, we measure
$\Delta\RM=0.76\pm0.59$~rad\,m$^{-2}$. The corresponding ratio gives
$\langle B_\parallel\rangle\approx240\pm180~\mu$G, about 50 times
below the value near closest approach in the same year. Part
of this drop is geometric: the impact parameter of the line of sight
grows from $5.1~R_\odot$ on 14 June to $23.8~R_\odot$ on 21 June, and
for a field falling as $r^{-2}$ this alone lowers $\langle
B_\parallel\rangle$ by a factor of about 20. The remaining factor, about 2, is not significant given the
uncertainty of the 21 June ratio.
Because 21 and 22 June are excluded from the DM baseline, they are also
excluded from the 2025 null test (Section~\ref{subsec:dmrm_method}).
Before this exclusion, 21 June alone set the null-test slope.
This behavior is not unique to 2025. In 2026 the post-conjunction tail
on 21--24 June, also excluded from the baseline, has
$\Delta\DM=(2.0$--$3.7)\times10^{-3}~\mathrm{pc\,cm^{-3}}$ but
$|\Delta\RM|\le0.7$~rad\,m$^{-2}$, giving $|\langle B_\parallel\rangle|\lesssim0.25$~mG
on each day. In both years the plasma sampled about a week after
conjunction carries a substantial column but little net
line-of-sight field.

Table~\ref{tab:bpar} lists $\langle B_\parallel\rangle$ for each year
and sector. The out-of-window null test is consistent with zero in
2022--2024 and 2026, and differs from zero by $2.3\sigma$ in 2025 ($+920\pm400~\mu$G), where the small residual $\Delta\DM$ of the
remaining post-conjunction days gives it some leverage. Off-window
$\Delta\DM$ is small, so the null test is weakly constrained in 2024 and
2026; it shows that the off-window days do not require a correlation,
not that their slope is smaller than the in-window values. The fits
with a free intercept return
intercepts consistent with zero, except for 10--13 June 2026
($-1.05\pm0.11$~rad\,m$^{-2}$). There the free-intercept slope
($\approx+740~\mu$G) is steeper than the through-origin value, which we
take as a sign that, as in 2025, the ratio is not constant across the
sector.
Where a single group spans the window, its reduced $\chi^2$ is 2.9 (2022). In 2025 it is 28: with DM uncertainties of
$0.2\times10^{-3}~\mathrm{pc\,cm^{-3}}$, a single ratio does not
describe the window. The daily ratio is $\approx11\,200~\mu$G on 13 and 14 June, falls to
$5400~\mu$G on 16 June, and is $1200$--$4800~\mu$G on 17--19 June, so the field sampled after
conjunction is weaker. The quoted 2025 uncertainty comes from the
day bootstrap and includes this spread. Fitting 2023, 2024 or 2026 as a single group instead gives reduced
$\chi^2$ of 7.2, 42 and 150, consistent
with those windows mixing opposite polarities, although the 2025 value
shows that a high $\chi^2$ can also arise without a sign change. The smallest statistical uncertainty, $+550\pm50~\mu$G, is for 10--13
June 2026 (four usable days), but the free-intercept fit above shows that the
ratio is not constant within that sector either.

\begin{deluxetable}{llrr}
\tablecaption{Density-weighted line-of-sight magnetic field near conjunction
\label{tab:bpar}}
\tablehead{
\colhead{Year} & \colhead{Days fitted} &
\colhead{$\langle B_\parallel\rangle$} & \colhead{Null test}\\
\colhead{} & \colhead{(June)} & \colhead{($\mu$G)} & \colhead{($\mu$G)}
}
\startdata
2022 & 10--20 & $+6200\pm1530$  & $-720\pm1920$   \\
2023 & 10--13 & $+1210\pm2380$  & $-500\pm3290$   \\
     & 14--20 & $-3490\pm670$   &                 \\
2024 & 12     & $\approx+810$\tablenotemark{a}  & $-1680\pm13620$ \\
     & 13     & $\approx-1130$\tablenotemark{a} &                 \\
     & 17--20 & $+1760\pm220$   &                 \\
2025 & 10--19 & $+9720\pm2200$  & $+920\pm400$    \\
2026 & 10--13 & $+550\pm50$     & $+550\pm3920$   \\
     & 14     & $\approx-1280$\tablenotemark{a} &   \\
     & 15     & $\approx+560$\tablenotemark{a}  &   \\
     & 16     & $\approx-290$\tablenotemark{a}  &   \\
     & 19--20 & $+240\pm60$     &                 \\
\enddata
\tablenotetext{a}{Single usable day; ratio $\Delta\RM/\Delta\DM$ rather than a fit.}
\tablecomments{Slopes of $\Delta\RM$ versus $\Delta\DM$ from orthogonal
distance regression through the origin, converted with
Equation~(\ref{eq:bpar}), rounded to 10~$\mu$G. Days within 10--20 June
are split at sign changes of $\Delta\RM$
(Section~\ref{subsec:dmrm_method}). The null test is the same fit applied
to days outside 10--20 June, excluding days removed from the DM baseline
(21--22 June 2025, 21--24 June 2026). The 2025 value is dominated by
14 and 16 June.}
\end{deluxetable}

\paragraph{Interpretation.}

Equation~(\ref{eq:bpar}) gives the line-of-sight field weighted by the
electron density near closest approach. Along a path several solar
radii from the Sun, contributions of opposite sign partly cancel, so
for the sectors whose slope differs significantly from zero,
$|\langle B_\parallel\rangle|$ is a lower limit on the typical field
strength (the 10--13 June 2023 sector is consistent with zero). For comparison, \citet{Ord2007} inferred $\gtrsim0.4$~mG along a line of
sight at $0.95^\circ$ elongation, closer to the Sun than the Crab's
minimum elongation of $\approx1.3^\circ$.
The sign changes within the 2023, 2024, and 2026 windows are naturally
explained by the line of sight crossing the heliospheric current
sheet, which separates regions of opposite magnetic polarity. The
largest $|\langle B_\parallel\rangle|$ occurs in 2025 and the smallest
in 2026. With five conjunctions we do not attempt to relate this to the
solar cycle.

Several limitations apply. The 2025 value and the 2023 negative sector
are each dominated by one or two days of large $\Delta\DM$. The near-zero exclusion of Section~\ref{subsec:rm_method} removes
observations whose RM before ionospheric correction lies within
$8.4~\mathrm{rad\,m^{-2}}$ of zero. With baselines of $-40$ to $-47~\mathrm{rad\,m^{-2}}$ and an
ionospheric RM of 1.7--5.6~rad\,m$^{-2}$, this corresponds to
$\Delta\RM\approx+26$ to $+53~\mathrm{rad\,m^{-2}}$. In 2022 and 2025 the measured values reach this range. The per-pulse
delay selection recovers 14 June 2022, but the 15 June 2025 observation,
between $+57$ on 14 June and $+24$ on 16 June, remains ambiguous and is
excluded from the fit even though its RM may be genuine. In 2024 two of the three sectors rest on a single day. The scattering index is fixed at
$\alpha=4$ in the DM fit; simulations show that a true index of 3.5 or
4.4 would change $\Delta\DM$ near conjunction by at most
$0.05\times10^{-3}~\mathrm{pc\,cm^{-3}}$ (Appendix~\ref{appendix:validation}, Table~\ref{tab:dm_validation}),
but the profile model does not capture every echo. Finally, the per-pulse selection recovers some, but not all, of the
days of largest $\Delta\DM$, which carry the most weight.

In 2025 the secondary DM enhancement has little associated RM. This
would be consistent with a density enhancement that carries either a
weak net line-of-sight field or a field that reverses within it. With
the present data we cannot distinguish these possibilities from a
non-solar origin of the DM feature.


\section{Summary and Outlook}
\label{sec:future}

We have measured the DM and RM of the Crab pulsar through five solar
conjunctions, 2022--2026, with CHIME. The DM rises toward every
conjunction and peaks on 14 or 15 June at
$(6.8$--$20.4)\times10^{-3}~\mathrm{pc\,cm^{-3}}$ above the
background. In 2025, and only in that year, a separate enhancement of
$(3.95\pm0.20)\times10^{-3}~\mathrm{pc\,cm^{-3}}$ follows on 21 June,
after the DM has returned to the background. RM excursions reach $+56~\mathrm{rad\,m^{-2}}$ and change sign within the window in three
of the five years, consistent with crossings of magnetic-sector
boundaries. The slope of $\Delta\RM$ against $\Delta\DM$ gives
$|\langle B_\parallel\rangle|$ of at least $\sim0.2$--10~mG for the
sectors with significant slopes, largest in 2025. The secondary 2025
DM feature and the post-conjunction DM tail of 2026 carry little RM.

A priority is to improve the profile model used for DM. Echoes in the
Crab Nebula can produce non-exponential scattering tails, and the
scattering index could be fitted using the full five-year data set
rather than fixed. Separate main-pulse and interpulse DMs and RMs would
provide an independent check of the uncertainties.
The per-pulse delay selection of Section~\ref{subsec:rm_method}
already recovers some of the days closest to the Sun. Pulse-by-pulse
polarization calibration, in particular of the leakage that we suspect
produces the near-zero RM solution, would allow the remaining ones to
be recovered.

Extending the DM and RM analysis to additional years will help
determine whether the two-feature structure observed in 2025 is
unusual or part of a broader pattern of year-to-year variability, and
why the 2025 peak was comparatively low. A comparison of
$\langle B_\parallel\rangle$ with potential-field source-surface models
\citep{Schatten1969,Altschuler1969} would test the sector
interpretation of the RM sign changes, by predicting the dates at
which the line of sight crosses the heliospheric current sheet in each
year, as \citet{Ord2007} did for a single pulsar. For DM, the
time-variable or Gaussian-process solar-wind models developed for
pulsar timing \citep{Tiburzi2021,Hazboun2022,Nitu2024,Susarla2024}
would provide a predictive baseline. Comparing that baseline with our
data would show whether the 2025 secondary enhancement departs from a
smoothly varying wind.

We also made an exploratory comparison with in situ measurements from
Parker Solar Probe \citep[PSP;][]{Fox2016PSP}, using Solar Probe Cup
(SPC) and SPAN-I data from the SWEAP suite
\citep{Kasper2016SWEAP}.\footnote{SWEAP data products, user guides, and
instrument release notes are available at
\url{https://sweap.cfa.harvard.edu/Data.html}.} For each June 2025
CHIME observation, PSP densities within $\pm10$ minutes set the
normalization of a spherically symmetric $n_e\propto r^{-2}$ profile,
which was integrated along the Earth--Crab line of sight. This
construction extrapolates a local measurement under a symmetry that
is known to underfit near-ecliptic lines of sight
(Section~\ref{sec:introduction}); the ion densities require a
composition assumption to give $n_e$; and contemporaneous sampling
does not establish that PSP and the line of sight intersect the same
plasma. SPC data were missing for 17--21 June, including the day of
the secondary enhancement, and the SPAN-I data did not establish an
association with it. The comparison therefore neither confirms nor
excludes a solar-wind origin of the 2025 feature. Extending it would
require documented data-quality selection, an explicit
spacecraft--line-of-sight geometry, and a background treatment
consistent with that of $\Delta\DM$. Coronagraph images and
coronal-mass-ejection catalogs could provide an additional test,
requiring consistency in propagation direction and travel time rather
than temporal proximity alone.


\appendix

\section{Validation of the DM and RM Estimators}
\label{appendix:validation}

We tested both estimators on simulated pulses with the same
channelization as the data and with flagged channels matching the data.

\paragraph{RM.}
We simulated 40 pulses with $\RM=-45~\mathrm{rad\,m^{-2}}$, a
cross-hand delay of 1.0~ns, a random intrinsic position angle for each
pulse, and a single-pulse SNR comparable to the
observations. Averaging the per-pulse coherence surfaces recovers
$\RM=-44.92\pm0.23~\mathrm{rad\,m^{-2}}$ and
$\tau_{\mathrm{xy}}=0.81\pm0.19$~ns, and every bootstrap resample
falls in the correct branch. The recovered $\tau_{\mathrm{xy}}$ is
within $1\sigma$ of the injected value; it is also close to the
0.75--0.90~ns found in the data, and with a single injection we cannot
exclude a weak pull of the estimator toward $\sim0.8$~ns. Averaging $Q$, $U$, and $V$ across pulses
before fitting instead converges to $-88.5~\mathrm{rad\,m^{-2}}$, with
$\tau_{\mathrm{xy}}$ on the grid boundary. This happens because the
random position angles cancel the coherent polarization.
This test combines the per-pulse fits by averaging their coherence
surfaces. The estimator of Section~\ref{subsec:rm_method} starts from
the same per-pulse fits but takes the dominant peak of the retained
$\RM_i$; a test of that estimator, including the pulse selection and
injected leakage, is left to a future revision.

\paragraph{DM.}
Each simulated observation contains 150 pulses. For every pulse and
channel we draw two complex polarization voltages whose variance
contains the pulse signal, and form $I=|X|^2+|Y|^2$, so that the noise
and self-noise statistics match the data. The flagged channels are
copied from a real observation. The pulse is a Gaussian of width
$10~\mu$s, delayed by the injected $\delta\DM$ and convolved with a
scattering tail $\tau(\nu)=\tau_{600}(\nu/600~\mathrm{MHz})^{-\alpha}$;
its fluence scales as $\nu^{-2.5}$. Each pulse has a random arrival
offset of up to $\pm51~\mu$s, and a band-integrated SNR drawn from a
log-normal distribution with median 15. After the ${\rm SNR}\geq8$ cut,
about 130 pulses remain. The simulated pulses then pass through exactly
the steps of Section~\ref{subsec:dm_method}: normalization, subbands,
alignment, the joint fit with $\alpha$ fixed at 4, and 100 bootstrap
resamplings (300 for the data; 100 suffice here because the bias is
estimated from the spread between realizations). We ran 10 independent noise realizations of each
configuration. The configurations span $\tau_{600}=0$--$70~\mu$s with
injected $\delta\DM=0$, 2, 4, and $7\times10^{-3}~\mathrm{pc\,cm^{-3}}$.
Three variants test the assumptions of the fit: a true index of 3.5 or
4.4, a tail with 30\% of the fluence in a second exponential five times
longer, and a fit window extended from $+640~\mu$s to $+1.28$~ms.

\begin{deluxetable}{rrrrrrr}
\tablecaption{Bias of the recovered residual DM on simulated observations
\label{tab:dm_validation}}
\tablehead{
\colhead{$\tau_{600}$} & \colhead{$\alpha=4$} & \colhead{$\alpha=3.5$} &
\colhead{$\alpha=4.4$} & \colhead{Two-comp.} & \colhead{Long window} &
\colhead{Cross-corr.}\\
\colhead{($\mu$s)} & \colhead{} & \colhead{} & \colhead{} & \colhead{tail} &
\colhead{} & \colhead{}
}
\startdata
10 & $-0.028$ & $-0.100$ & $+0.032$ & $-0.150$ & $-0.028$ & $+0.945$ \\
20 & $-0.031$ & $-0.140$ & $+0.064$ & $-0.113$ & $-0.030$ & $+1.174$ \\
30 & $-0.030$ & $-0.165$ & $+0.080$ & $-0.096$ & $-0.031$ & $+1.291$ \\
45 & $-0.031$ & $-0.188$ & $+0.094$ & $-0.081$ & $-0.032$ & $+1.428$ \\
60 & $-0.031$ & $-0.208$ & $+0.104$ & $-0.075$ & $-0.030$ & $+1.499$ \\
70 & $-0.034$ & $-0.214$ & $+0.111$ & $-0.072$ & $-0.035$ & $+1.553$ \\
\hline
$70-30$ & $-0.004$ & $-0.049$ & $+0.031$ & $+0.024$ & $-0.004$ & $+0.262$ \\
\enddata
\tablecomments{Mean recovered minus injected $\delta\DM$, in
$10^{-3}~\mathrm{pc\,cm^{-3}}$, over 10 noise realizations, for an
injected $\delta\DM=4\times10^{-3}~\mathrm{pc\,cm^{-3}}$. The fit always
fixes $\alpha=4$; the column headings give the true index of the
simulated scattering. ``Two-comp.\ tail'' has 30\% of the fluence in a
second exponential five times longer; ``Long window'' extends the fit to
$+1.28$~ms; ``Cross-corr.'' is profile cross-correlation for the
$\alpha=4$ case. The last row is the change between
$\tau_{600}=30$ and $70~\mu$s, which is what remains after the
off-window background subtraction. The standard error of each entry is
$(0.003$--$0.008)\times10^{-3}~\mathrm{pc\,cm^{-3}}$.}
\end{deluxetable}

Table~\ref{tab:dm_validation} gives the results. With the correct index, the bias is between $-0.06$ and
$0\times10^{-3}~\mathrm{pc\,cm^{-3}}$ for $\tau_{600}\geq10~\mu$s,
0.5--1.5\% of the injected value, and it does not depend on $\tau_{600}$.
For unscattered pulses it reaches $-0.14\times10^{-3}~\mathrm{pc\,cm^{-3}}$
at the largest injected $\delta\DM$. A wrong index or a two-component
tail produces an absolute bias of up to $0.21\times10^{-3}~\mathrm{pc\,cm^{-3}}$.
This bias varies slowly with $\tau_{600}$, so it largely cancels in
$\Delta\DM$: between $\tau_{600}=30$ and $70~\mu$s, the range spanned by
the 2025 off-window days and the days near conjunction, it changes by at
most $0.05\times10^{-3}~\mathrm{pc\,cm^{-3}}$ (last row of
Table~\ref{tab:dm_validation}). Extending the fit
window changes the result by less than
$0.005\times10^{-3}~\mathrm{pc\,cm^{-3}}$, so truncating the tail at
$+640~\mu$s does not bias the fit. The scatter of the recovered values
between realizations is 1.2 times the median bootstrap uncertainty, so
the bootstrap uncertainties should be scaled up by about 20\%; this
is small compared with the excess term of Equation~(\ref{eq:dm_sys}).

An earlier version of the estimator did not apply the per-pulse SNR
selection. Pulses too weak to align returned $\delta\DM\approx0$ for an
injected value of $+3\times10^{-3}~\mathrm{pc\,cm^{-3}}$, with an
apparently small uncertainty. With the selection, such observations are
rejected rather than reported.


\begin{acknowledgments}

We thank the CHIME/FRB Collaboration and the teams responsible for
the CHIME observations and data products used in this work.
We thank Marten H.\ van Kerkwijk for helpful discussions.

U.P.\ is supported by the Natural Sciences and Engineering Research
Council of Canada (NSERC) [funding reference numbers RGPIN-2019-06770,
ALLRP 586559-23, and RGPIN-2025-06396], the Ontario Research Fund (ORF-RE
Fund), AMD AI Quantum Astro, the Academia Sinica [Grand Challenge Seed
Program AS-GCS-114-M02 and Investigator Award AS-IV-115-M04], and the
National Science and Technology Council (NSTC) of Taiwan [grant
no.\ 115-2123-M-001-009].

We acknowledge that CHIME is located on the traditional, ancestral,
and unceded territory of the Syilx Okanagan people. We are grateful to the staff of the Dominion Radio
Astrophysical Observatory, which is operated by the National Research
Council of Canada. CHIME operations are funded by a grant from the
NSERC Alliance Program and by support from McGill University,
University of British Columbia, and University of Toronto. CHIME was
funded by a grant from the Canada Foundation for Innovation (CFI) 2012
Leading Edge Fund (Project 31170) and by contributions from the
provinces of British Columbia, Qu\'ebec and Ontario. The CHIME/FRB
Project was funded by a grant from the CFI 2015 Innovation Fund
(Project 33213) and by contributions from the provinces of British
Columbia and Qu\'ebec, and by the Dunlap Institute for Astronomy and
Astrophysics at the University of Toronto. Additional support was
provided by the Canadian Institute for Advanced Research (CIFAR), the
Trottier Space Institute at McGill University, and the University of
British Columbia. The CHIME/FRB baseband recording system is funded in
part by a CFI John R. Evans Leaders Fund award to I.H.S.

We acknowledge the NASA Parker Solar Probe mission and the SWEAP team
for the use of their data.

Computations were performed on the Trillium supercomputer at the
SciNet HPC Consortium \citep{Loken2010,Ponce2019}. SciNet is funded by
Innovation, Science and Economic Development Canada; the Digital
Research Alliance of Canada; the Ontario Research Fund: Research
Excellence; and the University of Toronto.

\end{acknowledgments}

\facilities{CHIME, Parker}

\software{Astropy, NumPy, SciPy, Matplotlib, pandas, h5py, spinifex}


\section*{Data Availability}

The daily DM and RM measurements, the ionospheric RM corrections, and
the per-pulse RM fits used in this work are available from the
corresponding author on reasonable request. The CHIME baseband data are
held by the CHIME/FRB Collaboration. Parker Solar Probe SWEAP data are
publicly available from the SWEAP data archive
(\url{https://sweap.cfa.harvard.edu/Data.html}). The UQRG global
ionospheric maps and the IGRF geomagnetic model used for the
ionospheric correction are publicly available.


\section*{AI Disclosure}

Generative artificial intelligence tools were used during the preparation
of this manuscript for assistance with language editing, LaTeX formatting,
and code development and debugging. All scientific analyses, interpretation
of results, and conclusions were performed and verified by the authors.
The authors reviewed and take full responsibility for the final content
of the manuscript.


\bibliography{references,crab_propagation_with_simard}
\bibliographystyle{aasjournalv7}

\end{document}